\documentclass[12pt,a4paper]{article}
\usepackage[cp866]{inputenc}
\usepackage[T2A]{fontenc}
\usepackage{amsfonts,amsmath,amssymb,amscd,euscript,epsfig}
\usepackage[russian,english]{babel}
\normalbaselines
\catcode `\@=11
\@addtoreset{equation}{section}

\def\section{\@startsection {section}{1}{\z@}{-3.5ex plus -1ex minus
     -.2ex}{2.3ex plus .2ex}{\normalsize\bf}}
\def\subsection{\@startsection{subsection}{2}{\z@}{-3.25ex plus -1ex
minus
 -.2ex}{1.5ex plus .2ex}{\normalsize\bf}}

\def\thebibliography#1{\section*{References}
\list
  {[\arabic{enumi}]}{\settowidth\labelwidth{[#1]}\leftmargin\labelwidth
  \advance\leftmargin\labelsep
  \usecounter{enumi}}
  \def\newblock{\hskip .11em plus .33em minus -.07em}
  \sloppy
  \sfcode`\.=1000\relax}
 
\catcode `\@=12
\newcommand{\be}{\begin{equation}\label}
\newcommand{\ee}{\end{equation}}
\newcommand{\prt}{\partial}
\newcommand{\p}{\prime}
\newcommand{\bib}{\bibitem}

\begin{document}

\begin{center}
{\bf QUATERNIONIC ANALYSIS AND THE ALGEBRODYNAMICS} 

\vskip0.3cm

{\bf Vladimir V. Kassandrov}

(Institute of Gravitation and Cosmology \\
of People's Friendship University of Russia \\
E-mail: vkassan@sci.pfu.edu.ru)

\end{center}

\noindent
We present the ``algebrodynamical'' approach to field-particle theory  
based on a nonlinear generalization of the Cauchy-Riemann conditions 
to  non-commutative algebras of quaternion-like type. For complex 
quaternions the theory is Lorentz invariant and naturally carries  
some gauge and twistor structures. Point- and string-like singularities  
are considered as particle-like formations; their electric charge  
is ``self-quantized''. A novel ``causal Minkowski geometry with 
additional phase'' is presented that is induced by the structure of 
biquaternion algebra. On its background self-consistent algebraic dynamics 
of singularities (``ensemble of dublicons'') is considered.

\vskip1cm

\noindent 
MSC Scheme: 83C60; 83E99; 30G35. PACS Numbers: 03.65. Fd; 03.50.-z; 02.30.-f.

\section{On the commutative and non-commutative analysis and the algebrodynamics}

History of discovery and investigation of exeptional algebras like quaternions or 
octonions, as well as of numerous attempts to apply them for ``explanation of the 
structure of the World'', is highly dramatic and and full of still unjustified 
hopes~\cite{Hamilton,Penrose}. Bibliography on applications of quaternions in 
theoretical and mathematical physics during only XX century runs to thousands of  
articles~\cite{Gsponer}. Considerable part of them is devoted to the problem of  
construction of quaternionic analysis which in respect of the richness of  
internal properties and applications can be comparable with complex analysis. 
However, in opinion of the majority of contemporary mathematicians, this problem 
has not get its solution till now~\cite{atiah}. 

Meanwhile, the {\it commutative} analysis, that is, the analysis for functions  
taking values in some associative commutative algebra of finite dimension 
$n \ge 2$ (not necessarily with division), has been constructed by G. Sheffers 
as far as at the end of XIX century~\cite{Sheff} in full analogy with the 
complex analysis. At present it is used, in particular, in the conception of 
{\it polynumbers} and related Finsler geometries developing in the works of 
D.G. Pavlov and his group~\cite{Pavlov,Garasko}. Generalization of this version of 
analysis to {\it superalgebras} has been realized in the works of V.S. Vladimirov 
and I.V. Volovich~\cite{Volovich}.

Principal distinction of non-commutative and commutative cases has been noted 
by A. Sudbery~\cite{sud}: non-commutativity obliterates the difference between 
an initial $q$ and ``conjugated'' $q^*$ elements of algebra, making it possible 
to express them through each other using only constant basic elements (``units'') of 
algebra. In particular, for the algebra of Hamilton's quaternions $\mathbb{Q}$ 
for any $q\in \mathbb{Q}$ one gets (see ~\cite{Penrose}, p.121):
\be{conjugate}
q^* \equiv -\frac{1}{2}(q+I*q*I+J*q*J+K*q*K),
\ee
where $I,J,K$ are the three ``imaginary'' units of the quaternion algebra. 
That is why the definition, in analogy with the complex case, of a ``quaternionic 
analytical''  (``quaternionic holomorphic'') function as that independent on the 
quaternionic conjugated  argument, appears here to be senseless.

On the other hand, natural definition of the ``right'' (``left'') derivative  
$F^\p (Z)$ of a quaternionic function $F(Z),~~F:{\mathbb Q}\mapsto{\mathbb Q}$:
\be{derquater}
F^\p = dF * dZ^{-1}~~~(F^\p=dZ^{-1}*dF)
\ee
is also unproductive, since the requirement of existence and uniqueness of the limit 
(\ref{derquater}) (that is, of its independence on the path of convergence 
to zero of the increment $dZ$ in the $\bf E^4$-space of the algebra $\mathbb {Q}$) 
leads to a considerably over-determined system of PDE's which appears to be 
compatible only for the trivial case of a linear function (for details see, e.g., 
~\cite{sud,kass1}). There exist also additional considerations which convince oneself 
in the difficulty of construction of quaternionic (and, generally, of 
non-commutative) analysis (see, e.g.,~\cite{ego}).

Nonetheless, numerous attempts to bypass these difficulties have been 
undertaken of which most known is the conception of Fueter~\cite{sud,ego,fuet}. 
In many articles conditions of ``quaternionic analyticity'' (or their 
{\it biquaternionic} extension) have been formally written down in the form 
of a linear system of equation of Maxwell-like type (together with correspondent 
wave equation as an expected generalization of the 2D Laplace equation of 
complex analysis). All these attempts, however, cannot, perhaps, be considered 
as a successive version of quaternionic analysis. As to the more complicated problem 
of construction of {\it non-associative} analysis, say, over the algebra of octonions, 
none approaches to its solution are seen till now at all (nevertheless, see~\cite
{kass1}, section 10).

Let us return now to the case of commutative analysis. Modern exposition 
of the above presented approach of Sheffers may be found, e.g., in the monograph
~\cite{Vishn}. Therein, instead of the definition of (invariant) derivative one exploits the  
requirement on the {\it differential} of a function of algebraic variable 
be represented in an invariant ``component-less'' form. This makes it possible to expand the 
approach to all the (finite-dimensional) associative commutative algebras $\mathbb{A}$  
including those with null divisors, in particular to the algebras of double and 
dual numbers.

Specifically, let $F(Z)$ be an $\mathbb{A}$-valued function  $F:\mathbb{A}\mapsto
\mathbb{A}$ of algebraic variable $Z \in \mathbb{A}$. Sheffers formulated 
condition of its {\it differentiability in $\mathbb{A}$} as that of proportionality 
of linear parts of the increments (i.e., of differentials) $dZ,dF$ of the 
independent variable and the function respectively:
\be{sheff}
dF = H(Z)*dZ ,
\ee
where $H\in\mathbb{A}$ and $(*)$ denotes the operation of multiplication in 
$\mathbb{A}$. For algebras with division condition (\ref{sheff}) is evidently 
equivalent to that of existence and ``path-independence'' of the ratio of 
increments, i.e. of the derivative $H(Z)=dF*dZ^{-1} \equiv F^\p (Z)$ and, in the 
particular case of the algebra of complex numbers $\mathbb{C}$, immediately 
leads to the Cauchy-Riemann equations. In general case linear PDE's connecting 
partial derivatives of the components of $F(Z)$ follow from (\ref{sheff}) 
after elimination of the components of $H(Z)$ and are completely analogous 
to the CR equations for the functions of complex variable. As a whole, the commutative 
analysis created by Sheffers in many aspects reproduces the 2D complex one, so that 
a wide class of $\mathbb{A}$-differentiable functions obeying condition 
(\ref{sheff}) and containing, in particular, arbitrary polynoms of 
$\mathbb{A}$-variable can be constructed. 

Nonetheless, the transition from commutative case to the non-commutative associative 
algebras of quaternion type seems rather fascinating since those algebras 
$\mathbb{A}$, unlike the commutative ones, possess a wide group of continuous 
symmetries represented by  {\it internal automorphisms} 
\be{intaut}
q \mapsto a*q*a^{-1}, ~~~a\in \mathbb{A}, \forall q \in \mathbb{A}, 
\ee
preserving the multiplication law in $\mathbb{A}$. For algebra 
$\mathbb{Q}$ the automorphism group is known to be 2:1 isomorphic to the group   
of 3D rotations $SO(3)$ so that {\it the exceptional group of Hamilton's 
quaternions may be treated as the algebra of the Euclidean physical space 
$\bf E^3$}. Its extension to the field of complex numbers 
 -- the algebra of {\it biquaternions} $\mathbb{B}$ -- makes it possible to  
ensure the transition to the 4D space-time and to write down all the basic 
equations of relativistic field theory in a very compact and beautiful form  
(see, e.g., ~\cite{Berezin}). Finally, the version of (bi)quaternionic analysis 
earlier suggested by the author and exposed in the article below, allows to 
obtain a nonlinear Lorentz-invariant generalization of the Cauchy-Riemann  
equations and to built only on this base a self-consistent field-particle theory -- 
the so called {\it algebrodynamics}. The article is devoted to  
presentation of this (nonlinear) version of non-commutative analysis and 
to its realization in the framework of the algebrodynamical approach.

\section{Quaternionic differentiability and conformal mappings}

Correct way to generalize the approach of Sheffers to quaternion-like algebras 
consists, perhaps, in the explicit account of the property of non-commutativity 
of the algebras like  $\mathbb{Q}$ in the very definition of a differentiable 
function of $\mathbb{Q}$-variable. Specifically, we note that in the right-hand  
part of the expression (\ref{sheff}) one finds an invariant 
$\mathbb{A}$-valued differential 1-form of the most general type which can be 
constructed using only operations in the algebra $\mathbb{A}$. According to these 
considerations, in the case of a non-commutative (but still associative) algebra 
$\mathbb{A}$, condition (\ref{sheff}) may be naturally modified into the following 
{\it condition of $\mathbb{A}$-differentiability of a function $F(Z)$} (see~\cite
{kass1,kass2} and references therein): 
\be{ncd}
dF = L(Z)*dZ*R(Z).
\ee
Here $L,R:\mathbb{A}\mapsto\mathbb{A}$ are two the so called {\it semi-derivative} 
functions of $F(Z)$, left and right respectively. For a given $F(Z)$ obeying (\ref{ncd}) they are defined non-uniquely, up to a transformation 
$L\rightarrow \alpha L,~R\rightarrow \alpha^{-1} R$ in which the function 
$\alpha(Z)$ takes values in the  {\it centre} (commutative subalgebra) of the 
algebra $\mathbb{A}$. Thus, according to the definition (\ref {ncd}), {\it the 
problem of determination of functions differentiable in a non-commutative 
associative algebra $\mathbb{A}$ is the problem of enumeration of all  
the triples of functions $\{F(Z),L(Z),R(Z)\}$ which satisfy the condition  
(\ref{ncd})} (up to the above mentioned $\alpha$-equivalence of the 
semi-derivatives). 

For commutative algebras condition (\ref{ncd}) reduces back to (\ref{sheff})  
where now the ``derivative'' $H(Z)$ is formed from ``semi-derivatives'' as  
$H(Z)= L(Z)*R(Z)$. On the other hand, if in the general non-commutative case 
one takes, say, $R(Z)=E$  (expecting the existence of the unity element $E$ 
in the algebra $\mathbb{A}$ considered), then he returns back to the condition 
(\ref{sheff}) with $H(Z)=L(Z)$. However, as it was already noticed, at least for 
the quaternion-like algebras condition (\ref{ncd}) is too rigid, since it can be 
satisfied only by linear functions of the form $F=A*Z+B$ with $A,B$ being some 
constant elements of algebra at study (see, e.g., ~\cite{sud,dev}).

In general case condition of $\mathbb{A}$-differentiability (\ref{ncd}) 
defines a wider class of functions. In particular, for the algebra of Hamilton's 
quaternions $\mathbb{Q}$ condition (\ref{ncd}) appears to be algebraically equivalent 
to the condition of {\it conformity} of the mapping $Z\mapsto F(Z)$ in the Euclidean 
space $\bf E^4$~\cite{kass3,kass5,kass4}. Indeed, taking the quaternionic 
norm $N^2(q)=q_0^2+q_1^2+q_2^2+q_3^2$ of the elements in left- and right-hand parts 
of the relation (\ref{ncd}) and using the property of {\it multiplicativity} of 
norms 
\be{multnorm}
N^2(p*q)=N^2(p)N^2(q),~~ \forall p,q \in {\mathbb Q},
\ee
one obtains:
\be{conform}
 {\overline {ds}}^2\equiv N^2(dF)=N^2(L*R)N^2(dZ)\equiv\Lambda(Z)ds^2, 
\ee
so that any $\mathbb{Q}$-differentiable function indeed defines 
some conformal mapping $ds\mapsto {\overline {ds}}$ in $\bf E^4$ with the scale factor
$\Lambda(Z)=N^2(L*R)$. Let us notice that in this respect condition 
(\ref{ncd}) can be also regarded as a natural generalization of the condition 
of complex holomorphy.

However, it is well known (the so called {\it Liouville theorem}, see, e.g.,
~\cite{Dubrovin}), that in the $\bf E^4$-space conformal mappings form a finite 
15-parametric group, in contrast to the infinite-dimensional group  
of conformal mappings on a complex plane realized by analytical functions 
of complex variable. Each of these conformal mappings in $\bf E^4$ corresponds  
to some $\mathbb{Q}$-differentiable function obeying condition (\ref{ncd}). 
Namely, for inversion $F(Z)=Z^{-1}$ one has $dF=-Z^{-1}*dZ*Z^{-1}$, i.e. an 
expression of the form  like (\ref{ncd}). Analogously, one can easily verify 
corresponding statement for other independent conformal mappings in $\bf E^4$: 
translations, rotations and dilatation  -- as well as for arbitrary their 
sequences. In other words, transformations defined by $\mathbb{Q}$-differentiable 
functions form the group isomorphic to the conformal group of $\bf E^4$.

Thus, for exceptional algebra with division $\mathbb{Q}$ the class of  
$\mathbb{Q}$-differentiable functions defined by the condition (\ref{ncd}) is      
again too narrow to be applied in fundamental physics. One cannot, say, hope to 
use these functions in the capacity of fundamental physical fields. 
It occurs that for this purpose one should pass to complexification of 
$\mathbb{Q}$, that is, -- to the {\it algebra of biquaternions $\mathbb{B}$}.

\section{Biquaternionic differentiability and the equation of $\mathbb{C}$-eikonal}

Below we restrict ourselves to the case of the full $N\times N$ matrix algebras 
$\mathbb{A}=Mat(N)$ over  $\mathbb{R}$ or $\mathbb{C}$ 
(when $N=2$, one has the isomorphism of the full matrix algebra $Mat(2,\mathbb{C})\cong 
\mathbb{B}$ to that of biquaternions). For equivalent of the quaternionic norm -- 
the {\it determinant} of the matrix of differentials $dF$ in the left-hand part of 
(\ref{ncd}) -- one gets:
\be{norm}
\det\Vert dF \Vert = \det\Vert L(Z)*R(Z)\Vert \det\Vert dZ\Vert \equiv 
\lambda(Z) \det\Vert dZ \Vert.
\ee
In the case when both matrices $L,R$ are invertible, so that $\lambda(Z)\neq 0$, 
condition (\ref{norm}) defines some conformal mapping with the scale factor 
$\lambda(Z)$ of the infinitesimal (complex or real indefinite) ``metric''
represented by determinant in (\ref{norm}). In particular, for the algebra 
$\mathbb{B}$ we deal with conformal mappings in the complexified Minkowski space 
$\mathbb{C}\bf M$. 

The most interesting case, however, seems to be that one when $\det L =0$ (or, 
analogously, $\det R = 0$); under this condition the scale factor $\lambda(Z)=0$, 
and the relation (\ref{norm}) defines a mapping of the full vector space of 
$\mathbb{A}$ into the subspace of its elements -- null divisors (into the 
complex ``null cone'' in the case of the algebra $\mathbb{B}$). Such mappings 
may be named {\it degenerate conformal mappings}. They constitute an important 
and wide class:  in the context of algebrodynamical theory presented further 
in the article just these mappings (and corresponding differentiable 
$\mathbb{A}$-functions)  are identified with physical fields. In particular, 
{\it under complexification of quaternions the class of differentiable 
functions and related mappings considerably extends}.

In the $N\times N$ matrix representation condition of differentiability (\ref{ncd}) 
in component notation takes the form ($A,B,...=1,...N$):
\be{matnot}
\nabla_{AB}F_{CD} = L_{CA}R_{BD}
\ee
where $\nabla_{AB}$ corresponds to the operator of derivation with respect to    
coordinates $Z^{AB}$. For some fixed pair of indices $C,D$ denoting  
$F_{CD}\equiv \Sigma, ~L_{CA}\equiv \phi_A, ~R_{BD}\equiv \psi_B$ one gets instead  
of (\ref{matnot}): 
\be{matnot2}
\nabla_{AB}~ \Sigma = \phi_A\psi_B. 
\ee
Determinant of the matrix of semi-derivatives in the right-hand part of the 
equation, by virtue of the factorized structure, is identically null. Consequently,  
one gets the equation: 
\be{eikonal1}
\det\Vert \nabla_{AB} \Sigma \Vert = 0, 
\ee
which is necessarily satisfied by any matrix component $F_{CD}\equiv \Sigma \in 
\mathbb{R}$ or $\mathbb{C}$ of any function $F(Z)$  differentiable in $\mathbb{A}$.

Equation (\ref {eikonal1}) represents itself a  {\it nonlinear} 
analog of the Laplace equation from complex analysis, and here {\it nonlinearity 
arises as a direct consequence of the account of non-commutativity of algebra} in  
the very definition of the $\mathbb{A}$-differentiable functions (\ref{ncd}). In the case 
of biquaternion algebra $\mathbb{B}$ equation (\ref{eikonal1}) is just the  
{\it equation of complex 4-eikonal}. Indeed, introducing for brevity  
the following notations for coordinates in matrix representation:
\be{coormat}
Z^{00}=u,~Z^{11}=v,~Z^{01}=w,~Z^{10}=p,
\ee
and computing the determinant (\ref{eikonal1}), we come to the equation:
\be{ceicon}
(\nabla_u\Sigma)(\nabla_v\Sigma) - (\nabla_w\Sigma) 
(\nabla_v\Sigma) = 0, 
\ee 
which in the (complex) Cartesian coordinates $z^0=(u+v)/2,~z^3=(u-v)/2,~
z^1=(w+p)/2,~z^2=i(w-p)/2$ takes the familiar form of the eikonal equation:
\be{fameikon}
(\frac{\prt\Sigma}{\prt z^0})^2 - (\frac{\prt\Sigma}
{\prt z^1})^2 - (\frac{\prt \Sigma}{\prt z^2})^2 - (\frac{\prt 
\Sigma}{\prt z^3})^2 = 0 . 
\ee

In accord with the results of our paper~\cite{eik} (see also~\cite{kasspavl}), 
{\it general solution} of the complex eikonal equation consists of 
two different classes  which both can be obtained in an algebraic way 
from some ``generating'' (complex analytical) function of the {\it projective 
twistor variable}. 

Specifically, let us choose, in expression (\ref{matnot}) 
for the 4-gradient of complex eikonal function, one of the 2-spinors, say, 
$\psi=\{\psi_B\}$ and define then the 2-spinor  $\tau=\{\tau^A\}$ {\it incident} 
to it in the sense the {\it Klein-Penrose correspondence}~\cite{Penrose2}
\be{inc}
\tau=Z\psi,~~\leftrightarrow~~\tau^A = Z^{AB} \psi_B.
\ee
A couple of spinors $\{\psi_B,\tau^A\}$ connected by the incidence relation 
(\ref{inc}) is called the {\it (projective) twistor} of complex Minkowski 
space $\mathbb{C}\bf M$. 

Indeed, equation (\ref{inc}) as well as the spinor $\psi$ itself in equation 
(\ref{matnot}) are defined up to a multiplication by a nonzero complex scalar; 
therefore, only {\it three complex ratios} of twistor components are essentially 
defined. Let, for example, the spinor component $\psi_0$ be not zero; 
then, making use of the projective equivalence, one can choose the twistor gauge 
of the form $\psi_0=1$ and get for the above ratios:
\be{twistcalibr}
\psi_1=G,~~\tau^0=wG+u,~\tau^1=vG+p 
\ee
where $\{u,v,w,p\}$ are complex coordinates $Z^{AB}$ in representation (\ref{coormat}).

Let us choose now an {\it arbitrary} function $\Pi$ of three complex arguments 
-- components of the projective twistor 
\be{generate}
\Pi(\psi_1,\tau^0,\tau^1)\equiv \Pi(G,wG+u,vG+p)
\ee
It easy to check that resolving equation $\Pi=0$ with respect to the unknown $G(u,v,w,p)$, 
one obtains some solution of the complex eikonal equation (CEE) -- solution of the I 
class. Further, resolving  equation $d\Pi / dG =0$ with respect to $G$ again and 
substituting the obtained solution into the initial function $\Pi$, we come to a 
``conjugated'' solution to CEE $\Pi(u,v,w,p)$ (of the II class). According to results 
of the paper~\cite{eik}, these two classes {\it exhaust all (almost everywhere  
analytical) solutions of the CEE} (see \cite{eik,kasspavl} for details). 
For further needs let us only mention that, for any generating (``World'') 
function $\Pi$, corresponding solution of the joint system $\Pi=0, ~~d\Pi/dG=0$ 
defines the structure of singular set $\Pi(u,v,w,p)=0$ -- the locus of branching 
points of the eikonal function $G$ ($\Pi$) itself and, correspondingly -- of poles 
of its 4-gradient. Resolution of this algebraic system makes it possible, sometimes 
in lack of an explicit expression for the eikonal function itself, to determine 
the structure of its singularities (which may be extremely complicated). This locus  
defines also the structure of singularities of associated gauge fields and of 
the field of effective curvature being extremely important in the algebrodynamical 
approach. Corresponding examples are presented in~\cite{kasspavl,gauss,kt,kr3} and 
below (section 8).

\section{Global symmetries and splitting of the equation of $\mathbb 
{A}$-differentiability}

Let us return now back to examine the conditions of $\mathbb{A}$-differentiability 
(\ref{ncd}) in general non-commutative case of the matrix algebra  
$Mat(N,\mathbb{C})$. It is easy to check that this fundamental relation 
preserves its form under the following transformations:
\be{simult}
Z \mapsto PZQ^{-1},~~F(Z)\mapsto SF(Z)T^{-1},~~
L(Z)\mapsto SL(Z)P^{-1},~~R(Z)\mapsto QR(Z)T^{-1},
\ee
where $P,Q,S,T$ are four constant invertible and, in general, distinct matrices 
$N\times N$ (here and below the symbol of matrix multiplication is omited  
for simplicity). Digressing from dilatations (generally, with different scale 
factors for coordinates $Z$ and functions $F(Z)$), we shall further on set the   
determinants of all matrices equal to unity so that $P,Q,S,T \in SL(2,\mathbb{C})$.   

In the particular case of equality of the entire matrices one gets the {\it internal 
automorphisms} of the algebra at study which leave invariant both the trace and the 
determinant of matrices. When $N=2$, i.e. in the case of biquaternion algebra 
$\mathbb{B}$, the determinant defines the structure of a bilinear 
$\mathbb{C}$-valued form:
\be{determin}
\det \Vert Z \Vert = (z^0)^2-(z^1)^2-(z^2)^2-(z^3)^2
\ee
Thus, in account of the invariance of the trace $z^0$, automorphisms represent 
themselves the rotations of 3-dimensional complex space $\mathbb{C}^3$; the 
automorphism group  $Aut(\mathbb {B})$ is 2:1 isomorphic to the group of 
complex rotations $SO(3,\mathbb{C})$. In general case ($N > 2$) 
automorphisms look like linear transformations which keep invariant the trace and 
the holomorphic Finsler-like ``metrical'' form of the $N$-th order defined by the  
structure of matrix determinant. 

For simplicity restricting below to the case $N=2$, let us consider 
general symmetries of the {\it conditions of biquaternionic differentiability} (\ref{simult}). 
The coordinate transformations 
\be{coortrans}
Z\mapsto PZQ^{-1}, ~~P,Q\in SL(2,\mathbb{C}), 
\ee
evidently represent themselves the 6-parametrical rotations of the full vector 
space of algebra $\mathbb{C}^4$ which leave invariant the holomorphic  
``metric'' (\ref{determin}). These transformations form the group 2:1 
isomorphic to the group $SO(4,\mathbb{C})$. By this, the law of transformations 
of the semi-derivatives $L(Z),R(Z)$ and the function $F(Z)$ itself remains, according to  
(\ref{simult}), partially indefinite due to existing voluntarism in the choice of 
two other matrices $S,T\in SL(2,\mathbb{C})$. This situation is, of course, 
related to a very wide symmetry group of the conditions of $\mathbb{B}$-differentiability 
(\ref{ncd}).

Indeed, one can set, in particular, $S=Q,~T=P$ considering thus symmetries of the form 
\be{symvector}
Z \mapsto PZQ^{-1},~~F(Z)\mapsto QF(Z)P^{-1}~~L(Z)\mapsto QL(Z)P^{-1},~~R(Z)\mapsto QR(Z)P^{-1},
\ee
under which all the ``fields'' $L(Z),R(Z),F(Z)$ behave themselves as (covariant) 
vectors realizing in this way vector representation of the group $SO(4,\mathbb{C})$.  
However, for the same fields another type of transformations preserving the form of 
basic equations 
(\ref{ncd}) is possible. Specifically, let us set the matrices $S,T$ equal to the unit 
one; then we come to the symmetry transformations of the form 
\be{symspinor}
Z \mapsto PZQ^{-1},~~F(Z)\mapsto F(Z),~~L(Z)\mapsto L(Z)P^{-1},~~R(Z)\mapsto QR(Z),
\ee
so that under these the principal function $F(Z)$ behaves itself as a $SO(4,\mathbb{C})$-scalar, 
whereas the semi-derivatives $L(Z),R(Z)$ --  as a complex of two independently 
transforming columns (rows), i.e. as the $SO(4,\mathbb{C})$-spinors! 

Thus, in the considered case one has a unique situation when one and the same 
``physical field'' can be transformed according to a number of independent 
representations of the ``complex Lorentz group'' $SO(4,\mathbb{C})$ manifesting  
itself at the same as a vector, a couple of spinors or a number of scalars. 
 
The most general symmetries (\ref{simult}) form (in the 4:1 ratio) 
the $12\mathbb{C}$-parametrical group $SO(4,\mathbb{C})\times SO(4,\mathbb{C})$  
which one can imagine himself as the product of {\it coordinate} and {\it internal} 
groups. However, in respect to the transformations of ``fields'', representation  
of the full group cannot be uniquely decomposed into representations of  
each of constituents.

Indeed, matrices $S,T$ can be uniquely represented in the form 
$S=\Lambda Q,~T=\Pi P$ through some new matrices $\Lambda,\Pi \in SL(2,\mathbb{C})$. 
By this, the field transformations under general symmetries (\ref{simult}) 
take the form:
\be{mixed}
Z \mapsto PZQ^{-1},~F\mapsto \Lambda (QFP^{-1}) \Pi^{-1},~~
L\mapsto \Lambda (QLP^{-1}),~~R\mapsto (QRP^{-1})\Pi^{-1},
\ee
and acquire the following natural interpretation: with respect to the group of 
``coordinate'' transformations $SO(4,\mathbb{C})_{coord}$ all of the fields 
$L(Z),R(Z),F(Z)$ are (covariant) vectors; at the same time, with respect to 
the internal ``isotopic'' group $SO(4,\mathbb{C})_{int}$ each of semi-derivatives 
$L(Z),R(Z)$ behave itself as a couple of {\it isospinors} whereas the basic field 
$F(Z)$ is an {\it isovector}. However, this interpretation though suitable is 
quite not the only possible one as we have seen above.

Let us notice also that the coordinate space $Z$ can be reduced to the space 
of {\it unitary} matrices (Hamilton's quaternions) or to the space of 
{\it Hermitian} matrices for which the above introduced rectilinear coordinates 
$z_\mu$ turn to be real and the invariant form (\ref{determin}) represents the 
Minkowski metric. By this, the requirement of preservation of the introduced  
condition (of unitary, Hermitian etc. structure) imposes restrictions on the 
admissible general symmetry transformations (\ref{simult}) so that the 
symmetry group reduces to a smaller one. All such situations including  
admissible transformations of ``fields'' (which generally remain complex-valued)
can be easily examined. In particular, on the Hermitian coordinate subspace 
the {\it algebrodynamical field theory} based on the conditions of 
$\mathbb{B}$-differentiability (\ref{ncd}) will be automatically 
Lorentz invariant. This case will be discussed in details below.

To conclude the discussion of symmetries, let us note that {\it linear} transformations 
(\ref{simult}) that contain the  $SO(4,\mathbb{C})$-rotations and dilatations 
do not exhaust the whole group of symmetries of the $\mathbb{B}$-differentiability  
conditions (\ref{ncd}) which are also evidently invariant under the $4\mathbb{C}$ 
translations as well as under the  {\it inversions} in this space, so that  
the full group of symmetries contains at least the 
$15\mathbb{C}$-parametrical group of conformal mappings on the 4D complex space  
equipped with holomorphic metric (\ref{determin}).

Now, in accord with the wide group of their symmetries, conditions of 
$\mathbb{B}$-differentiability admit various forms of ``splitting'', i.e. of 
their reduction to simpler systems of equations. By this, of course, symmetry group 
of the reduced system will be smaller than the initial one. The most important 
example of the procedure is the row (column) splitting of the matrix of the basic 
function $F(Z)$~\cite{kass2}. 

Specifically, let us denote the two {\it columns} of this matrix as 
$F=\{\eta_1,\eta_2\}$ and the columns of the right semi-derivative as   
$R=\{\xi_1,\xi_2\}$. Тhen one reduces the initial matrix system to that  
of the following two equations of identical type:
\be{reduc}
d\eta_a =\Phi * dZ * \xi_a, ~~~a=1,2
\ee
and each solution of the full system may be built as an arbitrary composition 
of some two solutions of systems like (\ref{reduc}) with the same matrix ``field''
$\Phi(Z)$ (the field of left semi-derivative). Reduced system (\ref{reduc}) is 
form-invariant, in particular, under the following transformations of variables:
\be{transf}
Z \mapsto QZP^{-1}, ~~\xi \mapsto P\xi, ~~  \Phi \mapsto P \Phi Q^{-1}, ~~~
\eta \mapsto P\eta, 
\ee
under which the quantities $\Phi(Z)$ transform as a complex 4-vector 
and the ``fields'' $\eta(Z)$ и $\xi(Z)$ -- as the $SL(2,\mathbb{C})$-spinors 
~\footnote{It is obvious that transformations (\ref{transf}) do not exhaust 
all the symmetries of a system of equations from (\ref{reduc}) 
which result from general symmetry transformations (\ref{simult})}.

Reduction of the full system of equations of $\mathbb{B}$-differentiability 
to a simpler system of the form (\ref{reduc}) for two spinors (basic and 
additional) and one 4-vector may be called the {\it spinor splitting}  
of the primary system of equations ({\ref{ncd}).

The main class of solutions of the full system ({\ref{ncd}) can in fact be 
restored from an arbitrary solution of only {\it one} of the spinor 
systems (\ref{reduc}).  For this, it is sufficient to nullify, say,  	
the spinors $\eta_2$ and $\xi_2$ or regard them as proportional to the  
initial spinors, i.e. to set: 
\be{propor}
\eta_2=k\eta_1,~~ \xi_2=k\xi_1
\ee
with arbitrary {\it constant} complex factor of proportionality 
$k\in \mathbb{C}$. By this, the right semi-derivative will represent  
a degenerate matrix, $\det R(Z) = 0$, and the principal matrix function 
will differ from a degenerate one by an arbitrary constant matrix $C$:~ 
$F(Z)=C+H(Z),~\det H(Z)=0$. We note that the factor of proportionality $k$ 
cannot depend on the coordinates $Z$ in a nontrivial way what may be easily  
proved in account of the identical form of the ``field'' $\Phi(Z)$ for 
both spinor systems (see~\cite{diss}). 

The {\it degenerate} case that corresponds to the {\it degenerate conformal mappings} 
(see section 3) is in general the only physically nontrivial one. Indeed, in the 
non-degenerate case correspondent to the canonical conformal mappings in 
$\mathbb{C}$ the field strengths of gauge fields associated with 
$\mathbb{B}$-differentiable functions identically turn to zero~\cite{kass1,kass2,jos}. 
On the other hand, when the matrix of, say, the right semi-derivative is degenerate, its 
two columns are proportional at a point and, by virtue of constancy of the 
factor $k$, -- globally. Thus, we have shown that physically 
nontrivial solutions of basic equations of $\mathbb{B}$-differentiability 
(\ref{ncd}) all correspond to degenerate matrices and 
can be all obtained from the solutions of the fundamental spinor system 
\be{reduct}
d\eta = \Phi dZ \xi
\ee
through the procedure of trivial completion of the spinors $\xi(Z),\eta(Z)$ up to  
the full matrices with zero determinant (by this, spinors can in addition be 
multiplied by an arbitrary complex number).

\section{General solution of fundamental spinor system}

As the complex eikonal equation (CEE) for individual components of the principal 
spinor $\eta(Z)$, general solution of fundamental spinor system (FSS) (ФСС) 
({\ref{reduct}) consists of two different classes and is obtained by analogy 
with general solution of the CEE itself (section 3). Here we shall only announce 
its structure (full proof will be given elsewhere).

\bigskip

\noindent
{\bf Solutions of FSS of the I class}

\bigskip

\noindent
Let us define the twistor of the space $\mathbb{C}^4$
\be{twistbase}
{\bf W} = \{\xi,\kappa\} \equiv \{\xi, Z\xi\},
\ee
built on a spinor $\xi(Z)$ which satisfies the FSS (\ref{reduct}) 
together with some corresponding functions $\eta(Z),\Phi(Z)$. Let three its 
{\it projective} components be functionally independent; then one may 
also consider as functionally independent all four its  components 
\be{comtwist}
\{\xi_A, ~~~\kappa^A = Z^{AB} \xi_B\}.
\ee
By this, it may be shown that components of the principal spinor, on the contrary, 
are functionally dependent and may be considered as dependent on coordinates only 
through the components of the twistor (\ref{comtwist}):
\be{implic}
\eta_A (Z) = \eta_A (\sigma), ~~\sigma(Z)=\sigma(\xi,\kappa)\equiv \sigma   
(\xi(Z),Z\xi(Z)).
\ee
The choice of generating function $\sigma(\xi,\kappa)$ as well as of the 
functional dependence on it of the components of principal spinor 
$\eta_A (\sigma)$ may be quite arbitrary (certainly, if one provides 
necessary smoothness conditions).

It appears that dependence on coordinates of the components of spinor $\xi_A(Z)$ 
can be by this determined from the solution of algebraic system of two equations 
of the form:
\be{algsys}
\frac{d\sigma}{d\xi_B}=0.  
\ee  
Substituting after this the solution $\xi(Z)$ into (\ref{implic}), one obtains 
expression of the principal spinor  $\eta(Z)$. By this, the ``field'' matrix 
$\Phi_{AB}$ is degenerate and equal to 
\be{leftfield}
\Phi_{AB}=\frac{d\eta_A}{d\sigma}\frac{\prt \sigma}{\prt \kappa^B}.
\ee
Thus, any differentiable function of twistor variable $\sigma(\xi,\kappa)$ 
gives rise to a class of equivalent (with respect to the functional dependence 
of the spinor components $\eta_A$) solutions to FSS. These solutions are 
in evident correspondence to the CEE solutions of the I class described in 
section 3.

\bigskip

\noindent
{\bf Solutions of FSS of the II class}

\bigskip

\noindent
Let now three projective twistor components (\ref{twistbase}) be functionally  
dependent; then, again with account of arbitrariness of the choice 
of the fourth component of {\it general} twistor, one may consider  
that there exist {\it two} functional constraints between its components 
(\ref{comtwist}) of the form
\be{bonds}
\Pi^{(D)}(\xi_A,\kappa^A)=\Pi^{(D)}(\xi_A, Z^{AB}\xi_B)=0,~~~(D)=1,2.
\ee
Resolving this system of algebraic equations, one can find the explicit form  
of the spinor $\xi_B(Z)$. Differentiating equations (\ref{bonds}) with respect to 
coordinates, one can show (for details see ~\cite{eik}) that the components 
of $\xi$ satisfy differential equations of the form
\be{potent}
\nabla_{AB} \xi_{C} =\left [ -\frac{\prt \Pi^{(D)}}{\prt \kappa^A} 
Q^{-1}_{(D)C}\right ] \xi_{B} \equiv \Psi_{CA}\xi_{B},
\ee
where the notation $\Psi_{CA}$ for quantities in square brackets is introduced 
and $Q^{-1}_{(D)C}$ -- for the matrix, inverse to 
\be{matr}
Q^{(D)C}:=\frac{d\Pi^{(D)}}{d\xi_{C}}.
\ee
In the invariant Pfaffian form system of equations (\ref{potent}) may be written down  
as follows:
\be{pfaff}
d\xi = \Psi dZ \xi.
\ee
Under identification of the principal and additional spinors 
$\eta(Z) \equiv \xi(Z)$ and the function $\Psi(Z)$ with  the ``field'' $\Phi(Z)$ (i.e. 
with left semi-derivative ``field''), 
this system is evidently itself a solution of FSS correspondent to generating 
twistor functions (\ref{bonds}).  

Actually, this case is of especial significance for further applications 
of biquaternionic analysis in algebrodynamical framework; in preceding articles 
the system (\ref{pfaff}) and corresponding full matrix system  
\be{gse}
dF = \Psi dZ F,
\ee
in which $F(Z)$ is a {\it degenerate} ($\det F=0$) biquaternionic field 
constructed by means of two proportional spinors $\xi(Z)$, has been called the 
{\it generating system of equations} (GSE). Indeed, as we shall see later on,
any solution of the GSE naturally gives rise to a solution of free equations 
of Maxwell, Yang-Mills and other fundamental (massless) equations of 
relativistic fields. We note that from mathematical point of view the GSE 
represents itself a special case of the $\mathbb{B}$-differentiability conditions  
under which the {\it right semi-derivative $R(Z)$ is identified with the 
principal biquaternionic ``field'' $F(Z)$}. 

Let us present now the {\it general form} of the FSS solutions of II class that corresponds  
to some arbitrary composition of the two generating twistor functions 
$\Pi^{(D)}(\xi_A,\kappa^A)$. From these, resolving the algebraic equations    
(\ref{bonds}) for the spinor $\xi(Z)$ and computing the quantities $\Psi(Z)$  
by means of formulas (\ref{potent}),(\ref{matr}), one obtains a complete 
solution to the GSE (\ref{pfaff}). By this, it turns out that the components 
of the principal spinor $\eta(Z)$ may be arbitrary (and, generally, different) 
functions of twistor components (\ref{comtwist}):
\be{gensoln}
\eta_A(Z)=\eta_A (\xi,\kappa)\equiv \eta_A(\xi(Z),Z\xi(Z)). 
\ee

Let us note also that by virtue of the constraints (\ref{bonds}) {\it 
only two of these twistor components are actually independent}. 
Finally, corresponding expression for the ``field'' $\Phi(Z)$ 
is obtained from the already found solution of GSE $\{\Psi(Z),\xi(Z)\}$  and 
arbitrarily chosen dependence of components of the principal spinor 
(\ref{gensoln}) in the following way:
\be{potfin}
\Phi(Z) = (M\Psi(Z)+N(E+Z\Psi(Z)), ~~~ M:=\Vert \frac{\prt \eta}{\prt \xi} \Vert, ~~~
N:=\Vert \frac{\prt \eta}{\prt \kappa} \Vert,
\ee
where $E$ represents again the $2\times 2$ unit matrix. As the result, 
one obtains that any pair of independent functions of twistor variable  
$\Pi^{(D)}({\bf W})\equiv\Pi^{(D)}(\xi_A,\kappa^A)$ gives rise to a class of 
equivalent (in respect of the arbitrariness of mutual dependence of the components 
of the principal spinor  $\eta_A({\bf W})$) solutions of the FSS. Certainly, this 
class corresponds to the II class of solutions to CEE described in section 3. 

Thus, we come to the general solution of FSS (\ref{reduct}). Indeed, since  
from the three projective components of principal twistor (\ref{twistbase}) 
either all three or only two are functionally independent~\footnote
{Statement that at least two components of a generic twistor are always independent  
is proved, for example, in~\cite{jos}}, any solution to FSS
belongs either to the first or to the second class. That is why any (almost 
everywhere analytical) solution to FSS may be obtained from some generating 
function of twistor variable (I class) or from a pair of such functions (II class) 
through the above described {\it purely algebraic} procedure. In compare with  
general solution to the complex eikonal equation described earlier in section 3
(in a fixed gauge) and in articles~\cite{eik,kasspavl},  in the case of FSS 
there exists an additional arbitrariness of the choice (of dependence on 
twistor variables) of the components of the principal spinor which may be either 
functionally connected (for the I class solutions) or independent (for solutions 
of the II class).

Such arbitrariness may be naturally eliminated if one chooses as fundamental 
the generating system of equations (\ref{pfaff}) or corresponding full-matrix
system (\ref{gse}). All solutions of the latter belong already to the 
second class of the FSS solutions and are completely determined by the choice 
of a pair of generating functions of twistor variable (\ref{bonds}) (or, under 
fixing of gauge for projective twistor (see below) -- even by a sole 
generating ``World'' function). Therefore, we proceed now to the detailed 
examination of properties and solutions of this universal system of equations.

\section{Biquaternionic differentiability and the gauge fields}
 
In the {\it algebrodynamics}, conditions of biquaternionic differentiability 
(\ref{ncd}) and, particularly, principal case of them -- the generating 
system of equations (\ref{pfaff}), (\ref{gse}) -- are considered as the 
unique primary equations of physical fields identified with differentiable 
$\mathbb{B}$-functions. By this, in order to guarantee the theory to be  
relativistic invariant, one has to restrict the complex coordinates $Z$ to the  
subspace of Hermitian matrices $Z\mapsto X=X^+$ with the Minkowski metric 
$\det X =(x^0)^2-(x^1)^2-(x^2)^2-(x^3)^2$.  The GSE (\ref{pfaff}) takes then 
the following form: 
\be{gseX}
d\xi = \Psi dX \xi
\ee
and preserves it (including the Hermitian structure of coordinate matrix) under  
the following symmetry transformations:
\be{symmet}
X\mapsto P^+XP,~~\xi\mapsto P^{-1}\xi,~~\Psi\mapsto P^{-1}\Psi(P^+)^{-1},
\ee 
where the quantities $\xi(Z)$ and $\Psi(Z)$ behave themselves as an 
$SL(2,\mathbb{C})$-spinor and a complex 4-vector respectively. Of course, there 
exists also a more general symmetry group (\ref{gseX}), namely, the {\it conformal} 
group of Minkowski space, and just this fact predetermines the existence of 
{\it twistor} structure introduced above. 

It should be noted, however, that the property of Hermitiance represents itself 
some superfluous requirement which is not motivated by the internal structure of  
initial algebra of biquaternions. In the last section we shall demonstrate in 
which way the structure of Minkowski space is actually {\it encoded} in the 
structure of the full vector space $\mathbb{C}^4$ of the $\mathbb{B}$-algebra. 
In account of this circumstance, in this and subsequent sections we preserve, as a 
rule, the {\it holomorphic} structure of theory dealing, as before, with 
complex coordinates $Z=\{z_\mu\}$ and, correspondingly, -- with GSE in its  
previous form (\ref{pfaff}), (\ref{gse}). When only the theory acquires an explicit 
physical interpretation, we accomplish transition to the real coordinates 
$\{x_\mu\}$ or, in other words, -- to the Hermitian matrix of Minkowski space 
coordinates $X=X^+$.  

Let us recall now that the GSE (\ref{pfaff}) is {\it over-determined} (8 
differential equations for 6 unknown functions). Therefore, some conditions  
of compatibility (integrability etc.) must be fulfilled that allow to obtain 
from (\ref{pfaff})  some restrictions on both the spinor $\xi(Z)$ and the vector  
field $\Psi(Z)$. However, before we start to consider these, it is necessary to 
examine the {\it gauge} nature of the field $\Psi(Z)$ that turns to be essentially 
distinct from generally accepted one. Let us also note that further in this and 
subsequent sections we follow mostly the exposition of the discussed questions 
presented in~\cite{jos,kass4}.

It is easy to see that the well-known from the field theory gauge 
$U(1)$-transformations of the form  
\be{caltrans}
\xi\mapsto \exp{i\alpha(X)}\xi,~~\Psi\mapsto \Psi-i\nabla \ln \alpha,~~\alpha \in \mathbb{R}
\ee
or their natural complexification, do not leave the GSE form-invariant. 
Nonetheless, in our papers~\cite{kass2,jos} it was shown that this system 
possesses the so called ``weak'' (or ``restricted'') gauge symmetry 
under which the gauge parameter $\alpha$ depends on coordinates implicitly, 
only through the components of the transformed spinor $\xi(Z)$ itself and 
the spinor $\kappa(Z)=Z\xi(Z)$ twistor-conjugated to it:
\be{calweak}
\alpha=\alpha({\bf W})=\alpha(\xi,\kappa)\equiv\alpha(\xi(Z),Z\xi(Z)).
\ee
Such transformations that correspond to the projective transformations of twistor 
components, form a group which is a (proper) subgroup of the full gauge group 
$\mathbb{C}$ (the latter being the complexification of $U(1)$)~\cite{jos}). 
By this, the quantities $\Psi(Z)$ transform gradient-wise, that is, behave 
themselves as the {\it potentials} of some gauge field. As we shall see below, 
this field may be naturally associated with (complexified) {\it electromagnetic}
field.

Indeed, the GSE (\ref{pfaff}) can be considered as  
{\it condition for the spinor $\xi(Z)$ be covariantly constant} 
(absolutely parallel) with respect to the $\mathbb{B}$-valued differential 1-form 
of effective connection:
\be{conn}
\Omega=\Psi dZ .
\ee
Interestingly, in the 4-vector representation $\mathbb{B}$-induced connection 
(\ref{conn}) gives rise to an affine connection of the form~\cite{kass1,kass2}: 
\be{weyl}
\Gamma_{\nu\rho}^\mu = \delta_\nu^\mu \Psi_\rho + \delta_\rho^\mu \Psi_\nu - 
\eta_{\rho\nu}\Psi^\mu - i\epsilon^\mu_{.\nu\rho\lambda}\Psi^\lambda, 
\ee
that defines actually the effective complex {\it Weyl-Cartan} geometry.  In such 
$\mathbb{B}$-induced geometry the non-metricity Weyl vector  and the vector  
of the pseudotrace of the skew-symmetric torsion are proportional to each other and are 
expressed both through the components of the principal gauge field 
$\Psi(Z)$~\footnote 
{Absolutely parallel fields in the framework of Weyl geometry free of torsion 
have been studied in~\cite{kr3}; their properties are closely related to the symmetries 
of Weyl manifolds~\cite{hall}. For real connections of such type relations 
between the non-metrical and torsion parts were the object of consideration 
in~\cite{Tod}}. 

Making now use of the definition (\ref{conn}), let us rewrite the initial GSE 
(\ref{pfaff}) in the form 
\be{ccv}
d \xi=\Omega \xi
\ee
Dynamics of the connection  $\Omega(Z)$  may be obtained  through external 
differentiation of  (\ref{ccv}) that results in the condition of integrability  
of the form
\be{compat}
R \xi \equiv (d \Omega - \Omega \wedge \Omega) \xi = 0,
\ee
where (in parentheses) a {\it curvature  2-form} $R$ appears. Since the spinor  
$\xi$ is not arbitrary but subject to (\ref{ccv}), conditions of   
integrability (\ref{compat}) do not result in the zero value of curvature 
~\footnote{At this point our approach considerably differs from that 
accepted in the works of Buchdahl~\cite{buch}, Penrose~\cite{prlast} or  
Plebanski~\cite{pleb} who conjectured that integrability conditions 
resembling (\ref{compat}) should be fulfilled for arbitrary spinor field 
(or for a wide class of solutions to the so called ``exact'' systems of 
field equations)}. Quite remarkably, instead of trivial ``zero curvature'' 
requirement, integrability conditions (\ref{compat}) result in  
the {\it self-duality} of curvature~\cite{kass1,kass2}.     
                           
In order to demonstrate this, let us note that for connection of the type 
({\ref{conn}) the curvature $R$ is of the following, rather special form: 
\be{curv}
R = (d \Psi - \Psi dZ \Psi) \wedge dZ \equiv \pi \wedge dZ ,
\ee
in which a novel $\mathbb{B}$-valued 1-form $\pi$ arises, with components 
\be{compon}
\pi_{AC}=\pi_{AC B D}dZ^{BD}=
(\nabla_{BD}\Psi_{AC}-\Psi_{AB}\Psi_{CD})dZ^{BD}.
\ee
Now the integrability conditions (\ref{compat}) take the form
$(\pi \wedge dZ) \xi =0$, or, in matrix representation: 
$$
\pi_{AC BD} dZ^{BD} \wedge dZ^{CE} \xi_{E} =0 .
$$
With account of the symmetry properties of 2-spinors from the last relation 
one obtains:
$$
\pi_{A~C(B}^{~~C}\xi_{E)} = 0,
$$
so that for any nontrivial solution $\xi(Z)$ one has:
\be{compi}
\pi_{A~CB}^{~~C}\equiv \nabla_{C B}\Psi_{A}^{~~C}+
\Psi_{BC}\Psi_{A}^{~~C} = 0.
\ee
  
Further, making use of the standard procedure and decomposing the curvature 
(\ref{curv}) into the self- and the antiself-dual parts one finds that equations 
(\ref{compi}) represent just the {\it conditions for the self-dual 
part of curvature to vanish}. By this, another antiself-dual its part $\bar R$ 
has the form:           
\be{antiself}
\bar R_{A~(BC)}^{~~D}=\nabla_{~(B}^{C}\Psi_{AC)} - 
\Psi_{~(B}^{C}\Psi_{AC)}
\ee
and satisfies the additional integrability conditions $\bar R \xi=0$ 
(later in the article we do not make use of these conditions). 

Thus, though the curvature 2-form (\ref{curv}) of the connection 1-form (\ref{conn}) 
is not (anti)self-dual by itself (i.e. (anti)self-dual in the ``strong'' sense), 
it necessarily becomes antiself-dual {\it on the solutions to GSE}. 
For this reason this property of the effective curvature of GSE 
has been called  {\it weak (anti)self-duality}~\cite{kr1}.

From physical viewpoint, expression (\ref{antiself}) defines the field strength  
of some matrix gauge field; in particular, its diagonal part 
\be{diagmax}
F_{BC}=\bar R_{A~~(BC)}^{~~A}=\nabla_{~(B}^{A}\Phi_{AC)}
\ee
corresponds to the strength of (complexified) electromagnetic field 
whereas the trace-tree part (\ref{antiself}) defines the strength of a complex 
field of the {\it Yang-Mills} type~\footnote{In fact, here introduced field is not 
exactly what is generally accepted as the Yang-Mills one with the gauge group 
$SL(2,\mathbb{C})$ if one takes in account the restricted (weak) gauge symmetry.
However, the form of gauge equations is completely identical to that generally  
accepted. Restrictions take place only with respect to the class of the 
admissible solutions and their transformations into each other under the action of 
the ``weak'' gauge group}. Indeed, in account of the {\it Bianchi identities}
\be{bianchi}
dR \equiv \Omega \wedge R - R \wedge \Omega, 
\ee
self-duality of curvature $R+iR^* = 0$ immediately implies the fulfillment 
of free Maxwell equations for diagonal (electromagnetic) part of the 2-form  
$F=Tr(R)=R_{A}^{~~A}$:
\be{maxeq}
dF^*=0= dF \equiv 0,
\ee
as well as of Yang-Mills equations for trace-free part of curvature form 
${\bf F}_{A}^{~~B}=R_{A}^{~~B}-\frac{1}{2}
F\delta_{A}^{~B}$.

By this, though the electromagnetic 2-form $F$ is, generally speaking, 
$\mathbb{C}$-valued, by virtue of its self-duality it is reduced to the real 
2-form $\tt{F}$ connected with $F$ in the following way:
\be{freal}
F=\tt{F}-i\tt{F}^*.
\ee
Certainly, for this form homogeneous Maxwell equations are satisfied too so that 
the number of independent degrees of freedom turns to be equal to that for the 
ordinary real electromagnetic field. In explicit form for $\mathbb {C}$-valued  
strengths of ``electric'' $\vec E$ and  ``magnetic'' $\vec H$ fields one has from   
(symmetric part of) the integrability conditions (\ref{compi}): 
\be{eh}
\vec E + i\vec H = 0,
\ee
from where one gets $\Im (\vec H) = \Re (\vec E)$, $\Im (\vec E) = - \Re (\vec H)$ so  
that a pair $\{\Re (\vec E), \Re (\vec H)\}$ represents the $\mathbb{R}$-valued   
electromagnetic field subject to Maxwell equations. In addition, from 
(the skew-symmetric part of)  equations (\ref{compi}) 
one obtains the following ``inhomogeneous Lorentz condition'' 
~\cite{kass1,kass2} for the $\mathbb{C}$-valued electromagnetic potentials  
$A_\mu \leftrightarrow \Phi_{AD}$:
\be{loren}
\prt_\mu A^\mu + 2 A_\mu A^\mu = 0,
\ee
which must also hold identically on the solutions of GSE. Certainly, 
condition (\ref{loren}) is not gauge invariant by itself, in the accepted  
``strong'' sense; nonetheless, it {\it is} invariant with respect to the 
``weak'' gauge transformations (\ref{calweak}), under the requirement that the transformed 
potentials (together with some corresponding spinor field $\xi(Z)$) 
really satisfy the GSE. 

As to the Yang-Mills fields, they can be here always expressed through the strengths 
of electromagnetic field and the spinor $\xi_{A}$ itself and, therefore, cannot be 
considered as independent. Note that separately the real and imaginary parts of 
the trace-less component of the curvature ${\bf F}_{A}^{~~B}$ will no longer 
satisfy free Yang-Mills equations by virtue of non-linearity of the latters. 
That is why here the Yang-Mills fields are {\it essentially complex-valued}. Other 
properties and peculiarities of the Yang-Mills fields arising in the framework of 
algebrodynamical approach can be found, say, in ~\cite{kass2}.

\section{Null shear-free congruences of rays associated with the GSE}

Let us now consider restrictions on the principal spinor $\xi_A$ arising under  
elimination of potentials of the gauge fields from the GSE (\ref{pfaff}). For this 
purpose, let us write out the given Pfaffian system of differential equations in  
components:
\be{pocomp}
\nabla_{BA} \xi_C = \Psi_{CB}\xi_A.
\ee
Multiplying the latter by the orthogonal spinor $\xi^A$ with account of 
skew-symmetry of the spinor norm $\xi_A\xi^A=0$ we get:
\be{bsk}
\xi^A\nabla_{BA} \xi_C = 0, 
\ee
i.e. the system of nonlinear equations for the components of the spinor $\xi(Z)$. 
Let us note that under restriction of complex coordinates to the Minkowski 
subspace $\bf M$, as a consequence of (\ref{bsk}), one 
obtains a (well known in the framework of GTR) system of equations for the 
principal spinor of the so called {\it shear-free null (geodesic) congruence}   
(SFC) of 4-dimensional rectilinear ``rays'':
\be{sfc}
\xi^B \xi^C\nabla_{AB} \xi_C=0. 
\ee
At this point we must warn the reader that here and below, in contrast to the  
generally accepted formalism, we make no difference between the primed   
and unprimed spinor indices under the restriction of coordinates to $\bf M$.
This is made to preserve as much as possible the notations specific for the full 
complex space and surely will not lead to any misunderstanding.    

In our articles~\cite{jos,Protvino} it was shown that initial system (\ref{bsk}) 
differs from the SFC system (\ref{sfc}) only in a more rigid fixing of the gauge 
of the principal spinor $\xi$, and is completely equivalent to the latter in 
what is related to the {\it ratio} of spinor's components. In particular, 
{\it general solution} of the SFC system (and, therefore, -- complete description  
of all such congruence on the background of the Minkowski space $\bf M$) is explicitly 
related to its twistor structure and represented by the famous {\it Kerr theorem}
~\cite{dks,Penrose2} in the form of implicit algebraic equation: 
\be{kerr2}
\Pi(\xi,\kappa)=\Pi(\xi,Z\xi)=0,
\ee
where $\Pi$ is an arbitrary {\it homogeneous} function of twistor arguments. From 
the constraint (\ref{kerr2}) the ratio of spinor components may be found which  
only is defined by the SFC system of equations ({\ref{sfc}). Analogously, the more 
rigid system of equations for the principal spinor of GSE (\ref{bsk}) has, as it has  
been shown earlier (section 4), general solution (\ref{bonds}) in the form of two 
equations that contain some arbitrary and independent twistor functions 
$\Pi^{(D)}(\xi,\kappa)$. From these equations now both spinor components 
$\xi(Z)$ can be defined altogether.

It is well known that, in order to draw geometrically a SFC on $\bf M$, one has 
to define, via the principal spinor $\xi(Z)$,  the field of a (real-valued) null 4-vector $k_\mu(X)$ tangent to the 
(rectilinear) rays of the congruence as follows:
\be{tang}
k_\mu(X)=\xi^+\sigma_\mu \xi, ~~~\sigma_\mu =\{E,\sigma_a\},
\ee
where $\{\sigma_a\},~a=1,2,3$ are the Pauli matrices and $E$ -- the unit 
$2\times 2$-matrix.  

Now, resolving the Klein-Penrose condition of spinor incidence (\ref{inc}) 
restricted to $\bf M$,
\be{incred}
\kappa=X\xi,
\ee
with respect to the space-time points $X$, one obtains~\cite{Penrose2} that twistor field  
$\{\xi,\kappa\}$ together with the SFC tangent vector $k_\mu$ {\it is transported in 
parallel along rectilinear null directions defined by the vector itself}. By this, 
as a parameter of transportation along the rays one may choose the time coordinate 
itself~\cite{kasspavl,pirt05,number}. 

Let us note now that for physical applications only {\it projective} components of 
the GSE twistor are of importance that are defined, say, by the ratio of spinor components 
 $\xi_1/\xi_0=G$ and equal then to 
\be{twistcon}
\kappa^0 = wG+u,~~\kappa^1=vG+p,
\ee
where $u,v,w,p$ are the complex matrix coordinates (\ref{coormat}) two of which 
($u,v$) become real under the restriction to {\bf M} whereas the two others ($w,p$)
become complex-conjugated. By this, both systems (\ref{bsk}) and (\ref{sfc}) 
for fundamental spinor field $G$ turn to be equivalent to a pair of PDE's of the 
following  form:
\be{gsol}
\nabla_w G=G\nabla_{u}G,~~~\nabla_{v}G=G\nabla_p G,
\ee 
General analytical solution of equations (\ref{gsol}) for function $G(X)$ 
follows now from its gauge invariant representation 
($\ref{bonds}$) in the form of a unique algebraic equation~\footnote{This may be 
compared with general solution of the complex eikonal equation of the II class, see section 3}
\be{rsol}
\Pi(G, \kappa^0, \kappa^1)= \Pi(G, wG+u, vG+p)=0,
\ee
that implicitly defines the function $G(X)$. Here $\Pi$ is an arbitrary  
holomorphic function of three complex twistor variables. Equation 
(\ref{rsol}) expresses itself the fact of functional dependence of the three 
components $\{G,\kappa^0,\kappa^1\}$ of the projective twistor $\bf W$ 
associated with solutions to GSE. For the SFC equations (\ref{sfc}) 
this equation is well known representing the Kerr theorem in a fixed gauge. 

Let us notice now that solutions of  (\ref{gsol}) are defined almost everywhere 
except the branching points of the $G(X)$-function that correspond to  
{\it multiple} roots  of the Kerr equation (\ref{rsol}) and are defined 
by the condition of the form:
\be{rsin}
P:=\frac{d\Pi}{dG}=0.
\ee
Multiplying now one by another the two equations (\ref{gsol}) one can verify once more 
the fact of fulfilment of the complex 4-eikonal equation for the field 
$G(X)$ in the form: 
\be{wafront}
\nabla_u G\nabla_v G -\nabla_w G \nabla_p G = 0,
\ee
On the other hand, differentiating these equations one can check that 
$G(X)$ satisfies also the linear {\it wave} (d'Alembert) equation
~\cite{kw,kr1,Protvino}
\be{wave}
\Box G \equiv (\nabla_u\nabla_v - \nabla_w\nabla_p)G = 0.
\ee
We mention also that in account of (\ref{wafront}) any $C^2$-function $\lambda(G)$
is also harmonic on the solutions of GSE:
\be{harm}
\Box \lambda(G) = 0.
\ee

Further, making use of the expression (\ref{potent}) for potentials $\Psi_{AB}$ 
and taking in account equation (\ref{wafront}), we can express the strengths 
of electromagnetic field (\ref{diagmax}) through the second order derivatives 
of $\ln G$: 
\be{streng}
F_{00}=\nabla_u\nabla_p\ln G,~~F_{11}=\nabla_v\nabla_w\ln G,~~
F_{01}=\nabla_w\nabla_p\ln G ,
\ee
so that fulfilment of free Maxwell equations for the strengths  
(\ref{streng}) follows directly from the wave equation (\ref{harm}) for 
$\lambda=\ln G$. Now, differentiating twice the identity (\ref{rsol}) 
with respect to the coordinates $\{u,v,w,p\}$, we obtain a very important  
(and having none analogues in literature) representation of the strengths of  
electromagnetic field (\ref{streng}) through the twistor variables
~\cite{jos,kass4}: 
\be{streng2}
F_{AB}=\frac{1}{P}\left(\Pi_{AB}-\frac{d}{dG}(\frac{\Pi_A\Pi_B}{P})\right),
\ee
where the function $P$ is defined by (\ref{rsin}) and $\{\Pi_A,\Pi_{AB}\},~ 
A,B=0,1$ denote the (first and second order) derivatives of the function $\Pi$ 
with respect to its twistor arguments $\kappa^0,\kappa^1$. Below we shall 
return back to this compact expression of the strengths of the associated 
electromagnetic field.

Close connections between the GSE and SFC equations gives us an opportunity                         
to introduce one more geometrophysical structure -- an effective {\it 
Riemannian metric}. Indeed, it is well known~\cite{dks,ksmh} that it is 
possible to deform the flat space-time metric $\eta_{\mu\nu}$ into a metric 
$g_{\mu\nu}$ of the {\it Kerr-Schild type}: 
\be{metr}
g_{\mu\nu} = \eta_{\mu\nu} + hk_\mu k_\nu 
\ee
so that all the defining characteristics of the SFC  -- geodesity, twist and 
shear-free property -- are preserved under such a deformation.
Here $h$ is some scalar function of coordinates, and the null (with respect to  
both the flat and deformed metrics) congruence $k(X)$ defined in 
(\ref{tang}) has the following projective invariant form: 
\begin{equation}\label{cong2}
k=du+\bar G dw + G d\bar w + G\bar G dv ,
\end{equation}
where as $\bar G$ the quantity complex conjugated to $G$ is denoted.

Let us turn now to the results of classical paper~\cite{dks} in which it 
has been proved that metric (\ref{metr}) satisfies the electrovacuum 
Einstein-Maxwell system of equations for functions $G$ obtained as the 
solutions of the Kerr algebraic equation (\ref{rsol}) with {\it linear} with  
respect to the twistor arguments $\kappa^0,\kappa^1$ generating functions $\Pi$~:
\be{PiG}
\Pi = \varphi + (qG + s)\kappa^1-(pG + \bar q)\kappa^0.
\ee
Here $\varphi=\varphi(G)$ is an arbitrary analytical function of the complex 
variable $G$, $s$ и $p$ are real and $q$ -- complex constants. Not going in 
details, we note that according to the results of paper~\cite{dks} 
scalar function $h$ in (\ref{metr}) is defined, up to an arbitrary constant, by   
initial generating function $\Pi$ and another function $\Psi(G)$ independent on 
$\varphi(G)$ and related to the electromagnetic field of the solution of Einstein-Maxwell 
system. Such fields are defined in the curved space with metric (\ref{metr}) and are, 
generally, different from those arising in our approach and satisfying Maxwell  
equations on the {\it flat} space-time background~\footnote{At the same time 
both these types of fields are, generally, different also from the fields 
which may be defined for any SFC through the twistor Penrose transform, 
see, e.g.,~\cite{Penrose2}, chapter 6}. 
Nonetheless, at least for the most physically interesting solutions (like those of 
Reissner-N\"ordstrem or of Kerr-Newman ones)  both these fields are  
nearly identical differing only in respect that in our approach electric 
charge is fixed in absolute value (due to the existence of ``master'' structure 
of the GSE (see section 8 below).

It was also shown in~\cite{dks,kw} that singularities of curvature of the 
effective Kerr-Schild metric (\ref{metr}) are defined just by the condition 
(\ref{rsin}). On the other hand, it follows from expression (\ref{streng2}) 
that the same equation $P=0$ defines the locus of singular points of 
associated electromagnetic field. The very same condition may be checked 
to define singularities of the Yang-Mills field associated with 
solutions of the GSE~\footnote{Additional singularities of the YM field strengths 
correspond to the {\it poles} of function $G(X)$~\cite{diss,jos}}. 

Thus, to any solution of the GSE it can be naturally put in correspondence 
some electromagnetic, complex YM and curvature (effective gravitational) 
fields. These satisfy respectively the free (complexified) equations of  
Maxwell, Yang-Mills and, at least in the basic stationary case -- the 
electrovacuum Einstein-Maxwell system~\footnote{Correspondence between   
shear-free null congruences and gauge fields has been studied for the case of a 
curved (algebraically special) space-time background in our paper~\cite{Trish2}}. 
Singularities of all these fields are defined by one and the same condition 
(\ref{rsin}) and completely coincide in space and time. This remarkable fact  
makes it possible, in the framework of algebrodynamical approach based on the GSE, 
{\it to consider particles as common singularities of all the associated fields}. 
We shall develop this conception in the subsequent section.

\section{Singular ``particle-like'' solutions of GSE with self-quantized electric charge}

We present here a brief review of the main classes of solutions of the GSE and 
of the associated Maxwell equations known for the present. All these can be  
obtained through the choice of a generating function $\Pi$, subsequent 
resolution of the algebraic Kerr equation (\ref{gsol}) and calculation of derivatives. If 
one restricts himself to the simplest case of solutions that can be obtained in 
{\it explicit} form, he has to consider only functions $\Pi$ {\it quadratic} in  
twistor arguments (linear functions lead to solutions with zero field strengths 
(\ref{streng}) of the associated electromagnetic field).  

Fundamental {\it static} solution is generated by the function $\Pi$ of the form  
\be{genfun}
\Pi = G\kappa^0-\kappa^1 + 2ia \equiv G(wG+u)-(vG+p)+ 2ia,
\ee
($a=Const \in \mathbb{R}$) which does not contain the time coordinate. Equating 
the function to zero and resolving the quadratic equation with respect to the  
unknown $G$ one gets (after restriction of coordinates to the real  
Minkowski space):
\be{kerr}
G = \frac{p}{(z+ia)\pm r_*}\equiv\frac{x+iy}{(z+ia)\pm \sqrt{x^2+y^2+(z+ia)^2}}.
\ee
Electromagnetic field (\ref{streng}) corresponding to the above solution, 
\be{elctrfd}
\vec E - i\vec H = \pm\frac{\vec r_*}{4(r_*)^3};~~~~~~~~(\vec E + i\vec H = 0),
\ee
where $\vec r_* =\{x,y,z+ia\}$, possesses the singular locus in the form of a {\it ring} 
of radius $a$, the only possible value of electric charge $q=\pm 1/4$ 
(in the dimensionless units used) and a dipole magnetic and quadruple electric 
moments equal respectively to $qa$ and $qa^2$~\cite{kr1,Protvino}. If one digresses 
from the restrictions on charge, the electromagnetic field (\ref{elctrfd}) together   
with the Riemannian metric (\ref{metr}) corresponding to the SFC (\ref{cong2}), 
precisely reproduces the field and metric of the Kerr-Newman solution 
(in the coordinates used in~\cite{dks}). In the particular case $a=0$  
solution (\ref{kerr}) corresponds to the {\it stereographic projection}  
$S^2\to \mathbb{C}$ and the fields turn into the Coulomb electric field and 
the Reissner-N\"ordstrem metric.

Self-quantization of electric charge is a fundamental property of the GSE
solutions discovered in~\cite{kass1,kass2}. This property follows from the 
self-duality conditions (\ref{eh}) which, together with the property of gauge 
invariance of the GSE, leads to restriction $q=N/4, N\in \mathbb{Z}$ for the 
admissible values of electric charge of electromagnetic field associated with any 
solution of the GSE~\footnote{Actually, in the $\mathbb{B}$-electrodynamics 
invariant with respect to the so called {\it duality transformations} it is not 
electric charge but the effective {\it magneto-electric} charge that is physically 
significant and quantized. In account of this, the problem of magnetic monopole 
also gets a natural solution~\cite{sing}}. Property of charge self-quantization has 
both topological and purely dynamical reasons,  
the latters being connected with the over-determined structure of the GSE. 
Proof of general theorem on charge quantization in the framework of algebrodynamics 
is presented in the articles~\cite{kass4,sing}.

By this, it is necessary to mention that, in contrast to some other,  
purely topological approaches to the problem of the charge quantization 
~\cite {ranada,zhuravl}, in the context of the GSE the charge of fundamental 
static solution (\ref{kerr}) can possess only a fixed and minimally posisble value   
and, consequently, can be naturally treated as  the {\it elementary charge}. 
Together with the known property of the Kerr-Newman solution 
to have the gyromagnetic ratio $g=2$, equal to that for the Dirac particle
~\cite{carter}, appearance of elementary charge in the theory justifies 
numerous attempts to interpret the ring singularity of fundamental 
solution (\ref{kerr}) in capacity of the classical model of electron. 
Such attempts have been undertaken, say, in the models of Lopes~\cite{lopes}, 
Israel~\cite{israel} or Burinskii~\cite{burin1} based exclusively on the 
properties of solutions of the Einstein-Maxwell system~\footnote{However, 
recently in our work~\cite{efimov} it has been proved that the Kerr congruence 
is {\it unstable} in the  sense that, under some small  
alteration of parameters of the generating function (\ref{genfun}),   
the static singular Kerr ring transforms into the ring uniformly expanding and then 
``irradiating to infinity''. Resolution of the instability problem requires,  
perhaps, a transition to the novel ``causal Minkowski geometry with phase'',   
see discussion in section 9 below}. 

According to general theorem proved in~\cite{kw} (see also~\cite{burin1}), 
all {\it static} solutions of the SFC equations (and, consequently, -- of the 
GSE) for which the singular locus is bound in 3D-space (below we call them  
{\it particle-like solutions}~\cite{kt}) reduce (up to 3D rotations and translations) 
to the Kerr solution (\ref{kerr}). If, however, 
one would remove requirement on a solution to be static and leave the  
class of functions (\ref{PiG}) considered in~\cite{dks}, he can find   
a lot of time-dependent ``particle-like'' solutions with bound singularities
of different dimensions, temporal dynamics and spatial shapes.

In particular, an {\it axisymmetric} solution of ``particle-like'' type 
generated by the function
\be{jozeph}
\Pi = \kappa^0\kappa^1 + b^2 G^2 = 0, ~~~b=Const,
\ee
has been found in~\cite{kr1,diss}. For real-valued $b$ it describes 
{\it two point-like singularities with elementary unlike charges $+1/4$ и $-1/4$} 
accomplishing a counter hyperbolic motion (i.e., uniformly accelerated).  
Electromagnetic field of this solution
\be{fieldborn}
E_\rho=\pm\frac{8b^2\rho z}{\Delta^{3/2}},~~~E_z = \mp\frac{4b^2 M}{\Delta^{3/2}},~~~
H_\varphi =\pm\frac{8b^2\rho t}{\Delta^{3/2}},
\ee
corresponds to that of the well known {\it Born solution}. By this, the following 
notations are used:
$$\rho^2 = x^2+y^2,~~~s^2 = t^2-z^2,~~~M=s^2+\rho^2+b^2,~~~
\Delta=M^2-4s^2\rho^2,
$$ 
and the field singularities are defined by the condition $\Delta=0$. 
For purely imaginary $b=ia,~ a\in \mathbb{R}$ initially, at $t=0$, one has an  
{\it electrically neutral} ring-like singularity of radius $a$ which in the course of  
time turns into an expanding {\it torus}. After the time passed 
$t = \vert a \vert$ singular locus transforms itself into a  {\it 
self-intersecting torus} represented at Fig.1.

\begin{center}
\begin{figure}[ht]
~~~~~~~~~~~~~~~~~
\epsfig{file=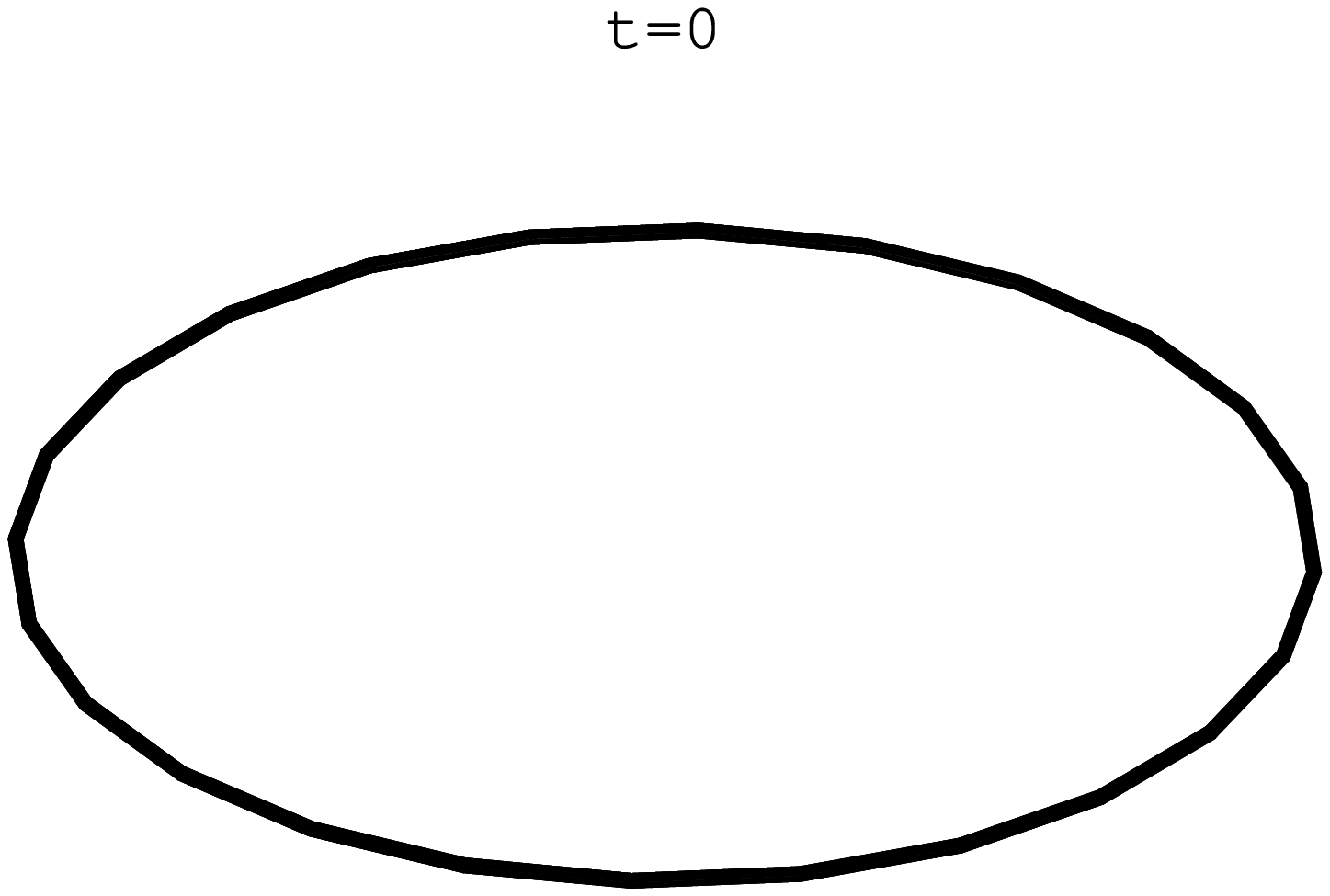, angle=-90, scale=0.3}
~~~~~~~~~~~~~~~~~~~~~~~~~
\epsfig{file=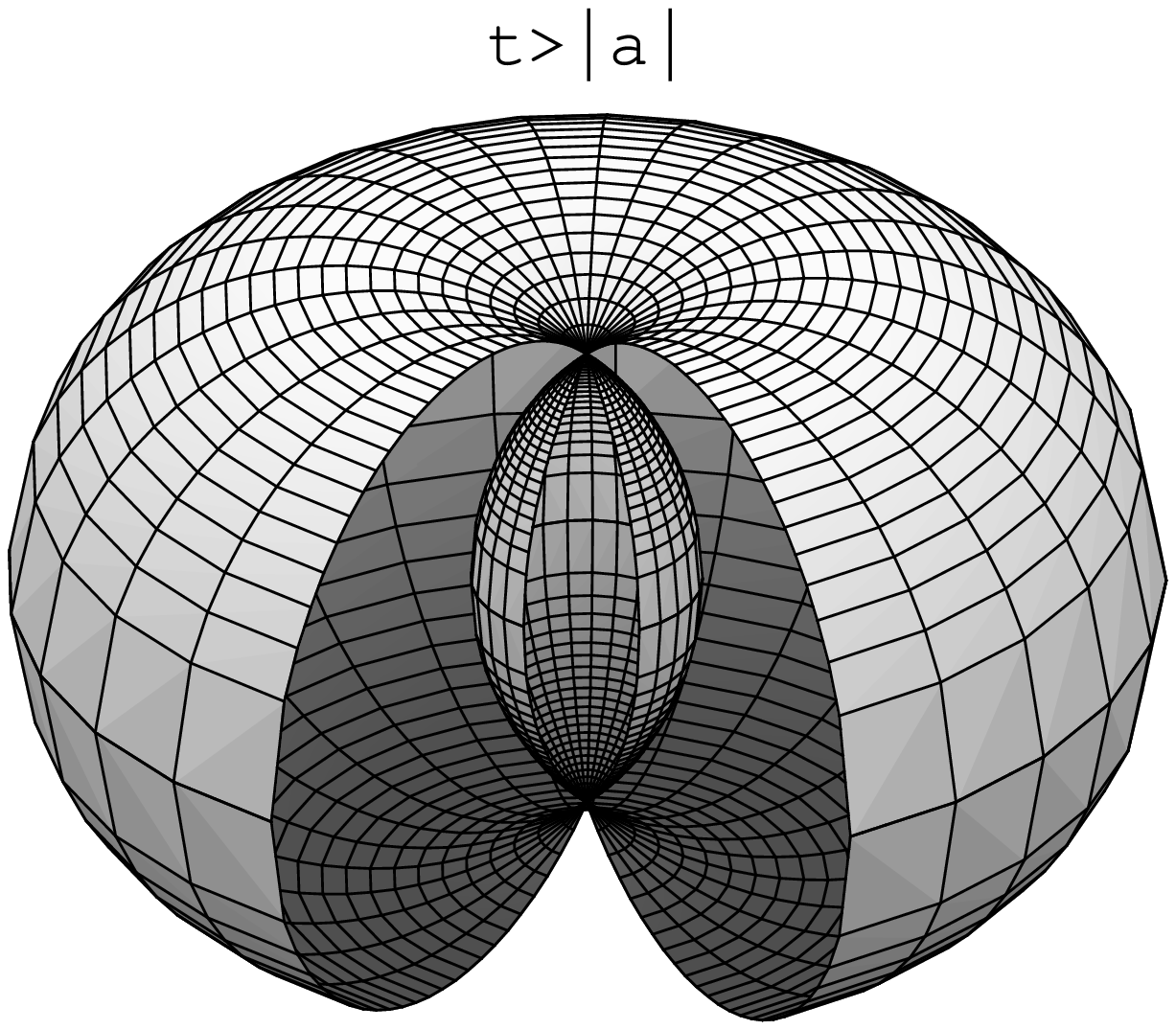, angle=-90, scale=0.4} 
\caption{Shape of the singular locus for electromagnetic field  
(\ref{fieldborn}) of electrically neutral solution (\ref{jozeph}) at initial  
($t=0$) and final ($t>\vert a \vert$) instants}   
\label{pic1}
\end{figure}
\end{center}

Let us also mention here a particle-like solution whose singular locus 
is {\it 8-shaped} (at initial moment), and a wave-like solution with {\it helix-like} 
singularity ~\cite{kt}. The latter (obtained from generating function 
more complicated than the quadratic one) represents itself an analogue of 
the electromagnetic wave in the algebrodynamical context.

If one gives up the condition for generating function to be quadratic, he    
comes to a wider class of the GSE solutions and to corresponding solutions of Maxwell    
equations with extremely complicated structure of singular locus. In particular, 
in~\cite{kasspavl,gauss} a solution of such type has been presented which  
describes the {\it process of annihilation} of a pair of oppositely charged 
point-like singularities. We have also found therein a solution with   
a ``photon-like'' singularity (in the form of a couple of crossed rings) 
moving uniformly and rectilinearly with the speed of light. 

Thus, in a purely algebraic way a wide class of explicitly or implicitly given 
solutions of {\it free} Maxwell equations with complicated and composite 
structure of point-like or extended singularities has been obtained. Considerable 
part of these solutions has not been known previously, and even their very 
existence has not been discussing. These solutions are well defined everywhere 
except the points at which the electromagnetic field strength turns to 
infinity. Locus of these singular points (at a given instant) may be 0-, 1- and 
even 2-dimensional (as it takes place for the case of the torus-like 
singularity (\ref{fieldborn})); moreover, it may dynamically change its 
dimension (say, for the same solution (\ref{fieldborn})). However, for 
solutions of {\it general type}, free of any type of symmetry, this singular  
locus is always one-dimensional and consists of a number of closed or 
infinite curves (``strings'')~\cite{gauss}. For solutions of {\it particle-like} 
type singular locus is bound in the 3D physical space.

Despite the initial ``vacuumness'' of gauge equations arising from the structure 
of GSE, the field singularities define spatial distribution and temporal   
dynamics of an effective {\it field source} at the points of whose location 
the property of analyticity of solutions becomes broken. Therefore, in contrast to 
the ordinary approach where an initially posed source defines its own electromagnetic 
field in the surrounding space, in here presented conception, on the contrary,  
{\it almost everywhere analytical field subject to free Maxwell (Yang-Mills) 
equations predetermines itself the location of its singular sources}. The 
considered solutions are well defined in the whole infinite space-time except 
at a singular set of zero measure. They are obtained algebraically from an 
arbitrary generating function and {\it do not require any initial or 
boundary conditions}. 

Moreover, it appears to be impossible, generally, to reduce this singular locus  
to a standard description through its covering by a family of $\delta$-shaped 
source distributions,  
owing to essential {\it multi-valuedness} of charged solutions of the Kerr type. 
Nonetheless, the whole set of ``quantum numbers'', the shape and dynamics 
of such singularities are correctly defined and quite nontrivial, this being   
related to the so called {\it hidden nonlinearity}~\cite{rantrub,ranada} 
of Maxwell (and Yang-Mills) equations in the framework of algebrodynamics, 
that is, to their {\it secondariness} with respect to the nonlinear structure  
of the primary GSE (as integrability conditions of the latter). 

It is just the presence of ``master'' equations -- the GSE -- that ensures the 
existence of a number of ``selection rules'' even for solutions of linear 
Maxwell equations, restrictions on admissible values of electric charge among  
them, and results also in the breakdown of the superposition principle 
(since, say, a sum of solutions satisfies the linear Maxwell equations but 
quite not necessarily -- the primary GSE itself!)

The over-determined primary GSE is, generally, not also invariant with respect    
to the spatial reflections (and, perhaps, -- to the time reversal)~\cite{kass2}. 
These invariances are restored only at the level of {\it consequences, integrability 
conditions} of these primary equations, namely -- at the level of Maxwell, 
Yang-Mills and similar equations.  Such situation seems to be exceptional in the 
field theory and, on the other hand, is completely adequate with respect to the 
observed physical reality. In seems to be even capable in principle to describe the 
P-violation and the time irreversibility.

More detailed discussion of the status of singular ``particle-like'' solutions 
in the algebrodynamics the reader can find in our works~\cite{kass2,kt,sing,kasspavl}.

\section{$\mathbb{B}$-induced complex space-time geometry and the ensemble of dublicons}

A beautiful representation of the solutions of SFC equations (and, 
consequently, -- of our biquaternion-induced GSE) has been suggested in the works 
of E.T. Newman et al.~\cite{newman,Lind,Newman2002} and developed later in 
the article~\cite{burin2} and in a series of subsequent works of A.Ya. Burinskii 
and  of E.T. Newman with collaborators. In this representation one considers a 
``virtual'' point-like charge ``moving'' along an arbitrary curve 
$\{z_\mu(\tau)\},~~\tau \in \mathbb{C}$ in the  {\it 
complexification} of Minkowski space $\mathbb{C}{\bf M}$. In this case 
the ``trace'' of the  {\it complex null (``light-like'') cone} of the 
``moving'' charge on the {\it real} Minkowski ``slice'' $\bf M$  of the full complex space 
forms there a null congruence of rays which appears always to be shear-free.

The Kerr congruence is only a simplest example of such representation (for this case, 
the generating point source is ``at rest'' at some point   
of ``imaginary'' (supplementary to $\bf M$) subspace of the full $\mathbb{C}\bf M$). 
The above presented solutions of the GSE and of corresponding SFC can all be obtained from 
such {\it Newman's representation}. On the other hand, these examples demonstrate  
that for such {\it ``complexified'' Lienard-Wiehert fields} the structure of  
singular locus can be very complicated and consists, generally, of a great 
number of one-dimensional curves -- strings.

In the algebrodynamical context complexified Minkowski space $\mathbb{C}{\bf M}$ 
arises unavoidably as the full vector space of the biquaternion algebra 
$\mathbb{B}$. At the same time, the above used procedure of restriction of 
coordinates to the real space-time $\bf M$ looks artificial and motivated only 
by physical considerations. Indeed, this subspace does not even form a subalgebra 
in $\mathbb{B}$ and is invariant neither under the $\mathbb{B}$-automorphisms nor 
under the full group of symmetry transformations (\ref{simult}). 

On the other hand, the group of $\mathbb{B}$-automorphisms $SO(3,\mathbb{C})$ 
consists of 6 real parameters and is 2:1 isomorphic to the Lorentz group $SO(3,1)$.  
One does not know any other group with properties like these. Quite reasonably,  
the algebra $\mathbb{B}$ and its symmetry group $SO(3,\mathbb{C})$  have been 
used in the works of A.P. Yefremov~\cite{Yefremov2} for construction of the 
so called {\it quaternionic theory of relativity} in context of which  
the invariant subspace $\mathbb{C}^3$ has been considered in capacity of the 
primordial space-time with three space-like and three time-like coordinates. 
In order to reduce the three-dimensional time to physical one-dimensional time, some   
additional requirements of orthogonality have been imposed. 

From the author's viewpoint, such ``exotic'' interpretation of the properties 
of biquaternion algebra is quite unnecessary. The matter is that its invariant 
subspace $\mathbb{C}^3$ can be in a natural way mapped into the ``causal'' domain  
of the physical Minkowski space-time equipped with additional internal 
fibre-like variables~\cite{kassmink}. Specifically, the principal invariant of 
initial complex space 
\be{Cinvar}
\sigma = (z_1)^2+(z_2)^2+(z_3)^2
\ee
can be separated into a non-compact ``modulus-like'' and compact ``phase-like'' parts. 
It is just the first part represented by the real-valued nonnegative invariant 
\be{modulinv}
S^2:=\sigma\sigma^*\ge 0,
\ee
that predetermines the observable ``spatially extended'' physical macro-geometry. 
whereas the second ``phase'' part of invariant $\sigma$ is perceived as 
defining the internal geometry of the ``fiber''. By this, the most important 
result of the above cited paper consists in the fact that the positively definite (or null)
$SO(3,\mathbb{C})$-invariant (\ref{modulinv}) can be identically 
represented in the form of Minkowski-like interval:
\be{minkcause}
S^2= \sigma\sigma^*\equiv T^2 - \vert \vec X \vert^2,
\ee
where the {\it real-valued} quantities 
\be{minkcoor}
T:=(\vec z \cdot \vec z^*), ~~~ \vec X:= i[\vec z \times \vec z^*]
\ee
under the $SO(3,\mathbb{C})$-rotations transform respectively as the time and space  
coordinates of the Minkowski space under the Lorentz-like transformations. In 
definition (\ref{minkcoor}) parentheses and square brackets denote 
the scalar and vector product of 3D (complex) vectors respectively. 

Thus, one can really consider the algebra of biquaternions $\mathbb{B}$ as the 
{\it space-time algebra}, and the Minkowski geometry is induced by this via   
the  {\it quadratic mapping} of complex coordinates of the invariant subspace 
$\mathbb{C}^3$ of the full vector space of $\mathbb{B}$ {\it into the internal, 
``causal'' domain of the light cone of $\bf M$} including its null boundary.
In this connection, apart of the {\it positively definite Minkowski interval (!)} 
(\ref{minkcause}) there arises another {\it phase invariant} of Lorentz  
tramsformations (precisely, of $SO(3,\mathbb{C})$-rotations) that might 
turn to be closely related to universal quantum properties of matter and to 
manifestations of quantum interference in particular.

According to the here discussed notions, the ``true'' primordial dynamics of  
matter-like formations (singularities, solitons etc.) takes place just in  
the initial complex space whereas the observable, ``shadow-like'' dynamics -- 
in the induced ``causal Minkowski space-time''. Such approach makes it 
possible, in particular, to successfully realize the beautiful old 
idea of Wheeler-Feynman about ``reproduction of electrons from one sole   
electron-germ''.

Specifically, let the point particle, in accord with the Newman's representation, 
``moves'' in $\mathbb{C}\bf M$ along a ``trajectory'' 
$\{z_\mu(\tau)\},~~\tau \in \mathbb{C}$ of general (sufficiently complicated) form. 
Then it can be shown~\cite{pirt05} that any position of this particle will be 
strictly correlated with other its positions on its own world line. Precisely, 
this correlation is established through equal values of fundamental twistor field 
of the null (complex) congruence generated by the particle-source and,  
correspondingly, -- through equation of the {\it complex null cone}.

The situation strongly resembles the known procedure for the {\it Lienard-Wiehert 
fields} in the framework of classical electrodynamics. However, in contrast to 
the case of real space-time, in complex space the ``light-like cone equation'' 
always have a considerable (if not infinite) countable number of roots. 
Consequently, a particle will ``see'' and ``receive signals'' ``from itself'' 
at different its locations on a unique trajectory. The arising set of 
identical yet differently located and moving particles has been named in our 
paper~\cite{pirt05} the ensemble of {\it duplicons}.

Apart of the idea of duplicons, the problem of {\it complex time} unavoidably  
arises in the context of complex dynamics. It turns to relate to general 
conception of physical time in the algebrodynamical paradigm~\cite{number,kasspavl,pirt05}. 
Specifically, to each of the GSE solutions there corresponds some shear-free 
null congruence of rays (section 7). This can be considered as the basic element 
of the pattern of the World arising in the algebrodynamics, namely, -- as the flow of 
primordial light, the so called {\it Prelight flow}~\cite{number,kasspavl}. 
In this connection, the whole ``matter'' represented in the theory by particle-like 
singular formations of associated fields appears as a set of {\it caustics} or 
{\it focal lines} of the fundamental Prelight flow.

Returning now back to the problem of time, let us note that on the real space-time 
$\bf M$ the time coordinate plays the role of the {\it parameter along the rays
of fundamental congruence} so that the defining property of time in this 
approach is the property of {\it reproduction}, of preservation of the primordial 
twistor field that takes place along the congruence rays (the ``rays of time'').   

Speaking figuratively, in the presented theory {\it time manifests itself as 
an automorphism of the primary field}. However, the electromagnetic and other associated 
fields expressible via the derivatives of fundamental twistor field are, of course, 
not preserved along the rays as well as the caustics-particles themselves.  
Just this circumstance defines another fundamental function of time that is related   
to the motion and {\it variability} of different forms of physical matter. 

Situation drastically changes in complex space $\mathbb{C}\bf M$ where the twistor  
field is defined up to a pair of arbitrary complex parameters and remains constant 
along the 2D {\it complex planes}~\cite{Penrose2,pirt05}. If, however, one 
requires in addition the property of preservation of the ``matter-like'' structure 
of {\it caustics}, then only {\it one} complex parameter remains free which 
thus can be interpreted as  {\it complex time}~\cite{pirt05}. Under this situation, 
however, there remains indefinite the {\it succession of the occurrences of events} 
(of the ``states'' of the Universe), and in absence of any grounds for its 
fixing it is the most natural to regard the alterations of complex time as 
{\it completely random} (casual). Then the arising for the ensemble of duplicons 
stochasticity can, apparently, be closely connected with quantum uncertainty and 
quantum theory in its Feynman's formulation in general. However, we are only 
coming to find the correct realization of these ideas.

\section{Conclusion}
 
In this article we did not regard as our principal goal to present a novel field model or a powerful  
algebraic method to obtain new complicated solutions of generally known  equations 
of classical field theory. Instead we here attempted to successively reveal the 
properties of the differentiable (analytical) functions of biquaternionic variable, 
that is, to develop a novel version of non-commutative analysis. Nonetheless, general 
conditions of $\mathbb{B}$-differentiability~\cite{kass1,kass2,kass3} reduce 
to the generating system of equations (\ref{pfaff}) which possesses 
an innate gauge and 2-spinor (twistor) structures and manifests remarkable connections 
with the structures and language generally accepted in the formalism of  
relativistic field theory. 

Essentially, it is sufficient to formulate only three principle 
conjectures in order to physically interpret the initially abstract mathematical
scheme:

\noindent 
1) on the space-time as a (real or invariant complex) subspace of vector 
space of $\mathbb{B}$-algebra, 

\noindent 
2) on physical fields as differentiable functions of $\mathbb{B}$-variable,   

\noindent 
3) on particles as (bound in 3D-space) singularities of strengths (curvatures) 
of the gauge and metrical fields directly associated with the primary 
$\mathbb{B}$-differentiable functions-fields.

From the physical viewpoint, the GSE may be considered as a rather specific 
system of field equations (nonlinear, non-Lagrangian, over-determined) for 
effectively coupled 2-spinor and electromagnetic (Yang-Mills) fields so that 
equations for both  are not postulated but follow directly from 
integrability conditions or ``contractions'' of the GSE itself.

Twistor structure also arises in a quite natural, ``dynamical'' way in the 
course of integration of the GSE and makes it possible to obtain the whole set 
of its solutions as well as those of equations for associated gauge fields in a 
perfectly simple algebraic way~\footnote{Note that in the Penrose's twistor 
approach~\cite{Penrose2,pr2} in order to obtain the solutions of wave-like 
(massless) equations it is sufficient to carry out an integration in 
twistor space; as to the presented scheme, even such integration is redundant  
therein}.

Particularly, from the algebraic Kerr equation (\ref{rsol}) a wide 
class of exact solutions of linear Maxwell equations with spatially 
extended yet bound structure of singularities can be directly obtained. 
In this connection, condition (\ref{rsin}) plays the role of the 
{\it equation of motion} for these particle-like formations and, at the same time,   
defines their characteristics (``quantum numbers'') and spatial distribution,
realizing thus the Einstein's conception of {\it super-causality}~\cite{einst}.

In consequence of tjhe breakdown of the superposition principle for solutions of 
``master'' equations -- the GSE -- temporal evolution of such particle-like 
objects simulates the process of physical interaction whereas  dynamical 
reconstructions ({\it bifurcations}) of the structure of singular locus can be  
treated as  {\it transmutations} of particles, in particular, as  
emission / absorption processes. All these processes obviously manifest  
close relationship to the theory of singularities of differentiable mappings and 
the catastrophe theory~\cite{argusein}.

We also hope that at least a number of remarkable properties of the GSE 
can be interesting in the general context of field theory. Let us note here, 
in particular: \\                                               
1) an opportunity to extend the class of physically important gauge field 
models  with account of the ``weak'' gauge symmetry (\ref{calweak}) discovered 
for the GSE or using the exceptional affine connections (\ref{conn}),(\ref{weyl}) of  
Weyl-Cartan type;\\
2) a natural opportunity to obtain proper ``selection rules'' for electric 
charge, spin and other physical characteristics starting from an over-determined 
system of field equations of the GSE type; \\  
3) complete algebraization of the primary PDE's for fundamental fields 
possessing twistor structure; \\
4) possibility to define the spatial distribution and the law of evolution of  
the field singularities without explicit resolution of the field equations themselves  
(but using instead the algebraic method of elimination of the principal field $G(X)$ 
from the joint system of equations (\ref{rsol}) и (\ref{rsin})).

One may imagine himself at least three possible points of view on the meaning of 
algebraic structures presented in this article and on the fundamental GSE in   
particular: as on a beautiful mathematical ``toy'', on a powerful method 
to obtain the solutions of the familiar field equations or, finally, as on a
fundamental dynamical system of equations primary with respect to generally  
accepted Lagrangian structures. In this connection, the construction 
of classical dynamics on the base of over-determined systems like GSE 
requires also quite new methods of quantization. On the other hand, at this point 
one can try to explain quantum properties as a whole via, say, the 
stochastic behaviour of an ensemble of particle-like field objects (dublicons,  
solitons, etc.) or invoking other yet purely classical and algebraic in nature 
methods and ideas.

In any case, in order to find a correct approach to quantization and 
to explanation of quantum properties of matter in general, it is necessary 
at the beginning to carefully study the properties of classical solutions 
on the background of ordinary Minkowski space-time and on the 
``phase extension of the Minkowski geometry" directly induced by the  
internal properties of the $\mathbb{B}$-algebra and briefly considered 
in the last section. We think  that just the underlying complex geometry 
can actually turn to be the true pregeometry of physical space-time and,  
moreover, be responsible for universal quantum properties of matter 
and quantum uncertainty in particular (in general context of an initially 
classical and deterministic theory).

To conclude, the already discovered properties of the $\mathbb{B}$-differentiable  
functions-fields and numerous geometrophysical structures they give rise to, 
looks like so unusually and, on the other hand, to such a great extent correlate 
with models and mathematical formalism of theoretical physics that force ourselves 
to ponder over possible {\it numerical} origin of fundamental laws of nature
~\cite{number,delphis} and to turn again, at the modern mathematical and physical 
level of comprehension, to the ancient philosophy of Pythagor, Plato and their 
followers.

\bigskip

\noindent
********************************************

\bigskip

The author is grateful to his disciples Jozeph A. Rizcallah and Vladimir N.  
Trishin for our long-term collaboration. Many important results presented 
in the article have been obtained at a time in our joint works. I am also 
indebted to Dmitriy G. Pavlov for his support and invitation to participate 
in the conference on Finsler geometry ``Cairo-2006''.

\end{document}